\documentstyle[11pt,amssymb,epsfig,cite]{article}

\newcommand{\labe}[1]{\label{equ:#1}}
\newcommand{\labf}[1]{\label{fig:#1}}
\newcommand{\refe}[1]{\ref{equ:#1}}
\newcommand{\reff}[1]{\ref{fig:#1}}
\newcommand{\Ref}[1]{Ref.~\cite{#1}}
\newcommand{\Eq}[1]{Eq.~(\refe{#1})}
\newcommand{\Eqss}[2]{Eqs.~(\refe{#1}) and (\refe{#2})}
\newcommand{\Eqssor}[2]{Eqs.~(\refe{#1}) or (\refe{#2})}
\newcommand{\Fig}[1]{Fig.~\reff{#1}}

\def\collab#1{{\bf #1\rm}}
\def\etalcollab#1{\etal\,(\collab{#1}),} 

\def\Bd{$B{^0_d}$}
\def\Bdbar{$\overline B{^0_d}$}
\def\Bs{$B{^0_s}$}
\def\Bsbar{$\overline B{^0_s}$}
\def\Bqq{$B{^0_q}$}
\def\Bqbar{$\overline B{^0_q}$}
\def\BdBdbar{\Bd--\Bdbar}
\def\BsBsbar{\Bs--\Bsbar}
\def\BqBqbar{\Bqq--\Bqbar}
\def\BBbar{$B{^0}\hbox{--}\overline B{^0}$}
\def\KKbar{$K{^0}\hbox{--}\overline K{^0}$}

\def\boldface{\bf} 
\def\bfit{\boldmath}
\def\ftstrut{\vrule height 5pt depth 0pt width 0pt} 
\def\textfrac#1#2{{\textstyle{#1 \over #2^{\ftstrut}}}} 
\def\EQN#1{\labe{#1}}
\let\DELTA=\Delta 
\def\IndexPageno#1{}
\def\lsim{\,\hbox{\char'056}\,} 
\def\etal{\hbox{\it et~al.}} 
\def\arns#1,#2(#3)
   {{\rm Ann.\ Rev.\ Nucl.\ Sci.\ }{\bf #1}, {\rm#2} {\rm(#3)}} 
\def\epjC#1,#2(#3){{\rm Eur.\ Phys.\ J.\ }{\bf C#1}, {\rm#2} {\rm(#3)}} 
\def\npB#1,#2(#3){{\rm Nucl.\ Phys.\ }{\bf B#1}, {\rm#2} {\rm(#3)}} 
\def\ptp#1,#2(#3){{\rm Prog.\ Theor.\ Phys.\ }{\bf #1}, {\rm#2} {\rm(#3)}} 
\def\plB#1,#2(#3){{\rm Phys.\ Lett.\ }{\bf B#1}, {\rm#2} {\rm(#3)}} 
\def\nim#1,#2(#3)
   {{\rm Nucl.\ Instrum.\ Methods\ }{\bf #1}, {\rm#2} {\rm(#3)}}
\def\prl#1,#2(#3){{\rm Phys.\ Rev.\ Lett.\ }{\bf #1}, {\rm#2} {\rm(#3)}} 
\def\prD#1,#2(#3){{\rm Phys.\ Rev.\ }{\bf D#1}, {\rm#2} {\rm(#3)}} 
\def\zpC#1,#2(#3){{\rm Z.~Phys.\ }{\bf C#1}, {\rm#2} {\rm(#3)}} 
\def\ijmpA#1,#2(#3)
   {{\rm Int.\ J.\ Mod.\ Phys.\ }{\bf A#1}, {\rm#2} {\rm(#3)}}
\def\npBps#1,#2(#3){{\rm Nucl.\ Phys.\ (Proc.\ Supp.) }{\bf B#1},
{\rm#2} {\rm(#3)}} 
\def\reference#1{\bibitem{#1}}

\begin{document}
\pagestyle{empty}
\begin{flushright}
IPHE 2002-009 \\
June 24, 2002 \\~
\end{flushright}

\vfill

\begin{center}

{\LARGE\bf \boldmath $B{^0}\hbox{--}\overline B{^0}$ MIXING}
\\[4ex] {\large O.~SCHNEIDER} \\[1ex]
{\it Institut de Physique des Hautes Energies} \\
{\it University of Lausanne, CH-1015 Lausanne} \\[1ex]
{\it E-mail:} {\tt Olivier.Schneider@cern.ch} \\ ~

\vfill

\begin{minipage}{0.8\textwidth}
The subject of particle-antiparticle mixing in the neutral $B$ meson systems 
is reviewed. The formalism of $B{^0}\hbox{--}\overline B{^0}$ mixing and 
basic Standard Model predictions are given, before experimental issues are 
discussed and the latest combinations of experimental results on 
mixing parameters are presented, including those on 
mixing-induced $CP$ violation, mass differences, and decay-width differences.
Finally, time-integrated mixing results are used to improve our knowledge on 
the fractions of the various $b$-hadron species produced at high energy 
colliders.
\end{minipage}

\vfill ~ \vfill

{\it To appear in the 2002 edition of the 
``Review of Particle Physics'',
} \\ {\it
K.~Hagiwara et al.\ (Particle Data Group),
Phys.\ Rev.\ D66, 010001 (2002).
}\\[2ex]
\end{center}

\newpage
~

\newpage
\pagestyle{plain}\setcounter{page}{1}   

\noindent
{\Large\bf \boldmath \BBbar\ MIXING}

\vspace{3mm}
\noindent Written March 2000 and revised March 2002 
by O. Schneider (University of Lausanne).

\section*{Formalism in quantum mechanics}

There are two neutral \BBbar\ meson systems, \BdBdbar\ and \BsBsbar\
(generically denoted \BqBqbar, $q=s,d$), which 
exhibit the phenomenon of particle-antiparticle mixing\cite{textbooks}.
Such a system is produced in one of its two possible states of 
well-defined flavor:
$| B{^0}\rangle$ ($\overline{b}q$) or $|\overline B{^0}\rangle$ ($b\overline{q}$).
Due to flavor-changing interactions, 
this initial state evolves into a time-dependent 
quantum superposition of the two flavor states, 
$a(t) | B{^0}\rangle + b(t) |\overline B{^0}\rangle$, 
satisfying the equation
\begin{equation} 
i \frac{\partial}{\partial t} 
 {a(t)\choose b(t)} = 
\left({\hbox{\boldface M}} - \textfrac{i}{2} {{\boldface\bfit \Gamma}}\right) 
 {a(t)\choose b(t)} \, ,
\end{equation}
where {\boldface M} and ${\boldface\bfit \Gamma}$, 
known as the mass and decay matrices, 
describe the dispersive and absorptive parts of \BBbar\ mixing.
These matrices are hermitian, and $CPT$ invariance requires 
$M_{11}=M_{22}\equiv M$ and $\Gamma_{11}=\Gamma_{22}\equiv \Gamma$ 

The two eigenstates of the effective Hamiltonian matrix 
$(\hbox{\boldface M} - \textfrac{i}{2} {\boldface\bfit \Gamma})$ 
are given by
\begin{equation} 
|B_{\pm}\rangle = p | B{^0}\rangle \pm q |\overline B{^0}\rangle \,,
\EQN{eigenstates}
\end{equation} 
and correspond to the eigenvalues
\begin{equation} 
\lambda_{\pm} = \left(M -\textfrac{i}{2} \Gamma \right)
\pm \frac{q}{p} \left(M_{12} -\textfrac{i}{2} \Gamma_{12} \right) \,,
\end{equation} 
where 
\begin{equation} 
\frac{q}{p} = 
\sqrt{\frac{M^*_{12}-\textfrac{i}{2}\Gamma^*_{12}}{M_{12}-\textfrac{i}{2}
\Gamma_{12}}}
\,. 
\EQN{alpha}
\end{equation} 
We choose a convention where ${\rm Re}(q/p) > 0$ and $CP | B{^0}\rangle = 
|\overline B{^0}\rangle$.

An alternative notation is
\begin{equation} 
|B_{\pm}\rangle = 
  \frac{(1+\epsilon) |B{^0}\rangle \pm 
        (1-\epsilon) |\overline B{^0}\rangle}{\sqrt{2(1+|\epsilon|^2)}}
~~~ \hbox{with} ~~~ 
\frac{1-\epsilon}{1+\epsilon} = \frac{q}{p} \,.
\end{equation} 

The time dependence of these eigenstates of well-defined masses 
$M_{\pm} = {\rm Re}(\lambda_{\pm})$ and decay widths 
$\Gamma_{\pm} = -2\,{\rm Im}(\lambda_{\pm})$ is given by the phases
$e^{-i\lambda_{\pm}t}= e^{-iM_{\pm}t}e^{-\textfrac{1}{2}\Gamma_{\pm}t}$:
the evolution of a pure $| B{^0}\rangle$ or
$|\overline B{^0}\rangle$ state at $t=0$ is thus given by
\begin{eqnarray}
| B{^0}(t)\rangle &=& g_+(t) \,| B{^0}\rangle
                     + \frac{q}{p} \, g_-(t) \,|\overline B{^0}\rangle \,,
\EQN{time_evol1}
\\
|\overline B{^0}(t)\rangle &=& g_+(t) \,|\overline B{^0}\rangle
                     + \frac{p}{q} g_-(t) \,| B{^0}\rangle \,,
\EQN{time_evol2}
\end{eqnarray}
where
\begin{equation} 
g_{\pm}(t) = \frac{1}{2} \left(e^{-i\lambda_+ t} \pm e^{-i\lambda_- t}  \right)
\,.
\end{equation} 
This means that the flavor states remain unchanged ($+$) or oscillate into each other ($-$)
with time-dependent probabilities proportional to
\begin{equation} 
\left| g_{\pm}(t)\right|^2 = \frac{e^{-\Gamma t}}{2}
\left[ \cosh\!\left(
\frac{\DELTA\Gamma}{2}\,t\right) \pm \cos(\DELTA m\,t)\right] \,, 
\EQN{cosh_cos}
\end{equation} 
where 
\begin{equation} 
\DELTA m = |M_+-M_-| \,, ~~~ \DELTA \Gamma = |\Gamma_+-\Gamma_-| \,. 
\end{equation} 
Time-integrated mixing probabilities are only well defined when considering 
decays to flavor-specific final states, {\it i.e.} final states 
$f$ such that the instantaneous decay amplitudes 
$A_{\overline f} = \langle \overline f | H| B{^0}\rangle$ 
and $\overline A_f = \langle f | H|\overline B{^0}\rangle$, 
where $H$ is the weak interaction Hamiltonian,  
are both zero.
Due to mixing, a produced $B{^0}$ can decay to the final state 
$\overline f$ (mixed event) in addition to the final state $f$ (unmixed event).
Restricting the sample to these two decay channels, the time-integrated
mixing probability is given by
\IndexPageno{Btimm}
\begin{equation}
\chi_f^{B{^0} \to \overline B{^0}} = \frac{
\int_0^{\infty} |\langle \overline f |H| B{^0}(t)\rangle |^2 dt}{
\int_0^{\infty} |\langle \overline f |H| B{^0}(t)\rangle |^2 dt +
\int_0^{\infty} |\langle           f |H| B{^0}(t)\rangle |^2 dt} 
= \frac{|\xi_f|^2 (x^2+y^2)}{|\xi_f|^2 (x^2+y^2)+2+x^2-y^2} \,,
\EQN{mixprob} 
\end{equation}
where we have defined 
$\xi_f = \frac{q}{p}\frac{\overline A_{\overline f}}{A_f}$ and
\begin{equation} 
x = \frac{\DELTA m}{\Gamma}  
\,, ~~~
y = \frac{\DELTA \Gamma}{2\Gamma} \,.
\EQN{x}
\end{equation} 
The mixing probability $\chi_f^{\overline B{^0} \to B{^0}}$ 
for the case of a produced $\overline B{^0}$ is
obtained by replacing $\xi_f$ with $1/\xi_f$ in \Eq{mixprob}. It is
different from $\chi_f^{B{^0} \to \overline B{^0}}$ if $|\xi_f|^2 \ne 1$, 
a condition reflecting non-invariance under the $CP$ transformation. 
$CP$ violation in decay amplitudes
is discussed elsewhere\cite{CP_review} and we assume 
$|\overline A_{\overline f}| = |A_f|$ from now on. The deviation of 
$|q/p|^2$ from 1, namely the quantity
\begin{equation} 
1 - \left|\frac{q}{p}\right|^2
= \frac{4\,{\rm Re}(\epsilon)}{1+|\epsilon|^2} 
+ {\cal O}\left(\left( 
\frac{{\rm Re}(\epsilon)}{1+|\epsilon|^2}\right)^2\right)\,,
\end{equation} 
describes $CP$ violation in $ B{^0}\hbox{--}\overline B{^0}$ mixing. 
As can be seen from \Eq{alpha}, 
this can occur only if $M_{12}\ne 0$, 
$\Gamma_{12}\ne 0$ and if the phase difference between
$M_{12}$ and $\Gamma_{12}$ is different from 0 or $\pi$. 

In the absence of $CP$ violation, $|q/p|^2 = 1$, ${\rm Re}(\epsilon)=0$, 
the mass eigenstates are also $CP$ eigenstates, 
\begin{equation} 
{CP} \, |B_{\pm}\rangle  = \pm |B_{\pm}\rangle \,, 
\end{equation} 
the phases $\varphi_{M_{12}} = {\rm arg}(M_{12})$ and 
$\varphi_{\Gamma_{12}} = {\rm arg}(\Gamma_{12})$ satisfy
\begin{equation} 
\sin(\varphi_{M_{12}}-\varphi_{\Gamma_{12}}) = 0 \,,
\end{equation} 
the mass and decay width differences reduce to 
\begin{equation} 
\DELTA m = 2\,|M_{12}| \,, ~~~ 
\DELTA \Gamma = 2\,|\Gamma_{12}| \,, 
\EQN{Dm,DG} 
\end{equation} 
and 
the time-integrated mixing probabilities 
$\chi_f^{B{^0} \to \overline B{^0}}$ and 
$\chi_f^{\overline B{^0} \to B{^0}}$ become both equal to
\begin{equation} 
\chi = \frac{x^2+y^2}{2(x^2+1)} \,. \EQN{chi}
\end{equation} 

\section*{Standard Model predictions and phenomenology}

In the Standard Model, the transitions \Bqq$\to$\Bqbar\ and \Bqbar$\to$\Bqq\ 
are due to the weak interaction.
They are described, 
at the lowest order, by box diagrams involving
two $W$~bosons and two up-type quarks (see \Fig{box}), 
as is the case for \KKbar\ mixing.
However, the long range 
interactions arising from intermediate virtual states are negligible 
for the neutral $B$ meson systems,
because the large $B$ mass is off the region of hadronic resonances. 
The calculation of the dispersive and 
absorptive parts of the box diagrams yields the following predictions 
for the off-diagonal element of the mass and decay matrices\cite{Buras84},

\begin{figure}\begin{center}
\vskip -1.0cm 
    \parindent = 0pt \leftskip = 0pt \rightskip = 0pt%
    \vskip .4in%
    \leavevmode%
    \centerline{%
\psfig{figure=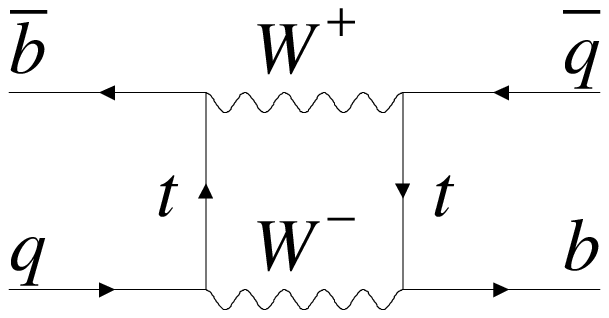,width=0.45\textwidth,clip=t}%
\psfig{figure=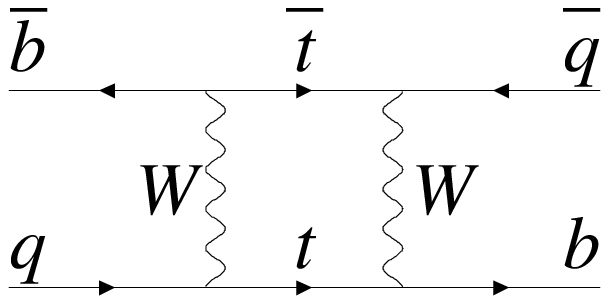,width=0.45\textwidth,clip=t}%
}%
    \nobreak%
    \vglue .1in%
    \nobreak%
\vglue -0.8cm
\caption{
Dominant box diagrams for the 
\Bqq$\to$\Bqbar\ transitions ($q = d$ or $s$). Similar 
diagrams exist where one or both $t$ quarks are 
replaced with $c$ or $u$ quarks.
} 
\labf{box}
\end{center}\end{figure}
\begin{eqnarray}
M_{12} &=& - \frac{
           G_F^2 m_W^2 \eta_B m_{B_q} B_{B_q} f_{B_q}^2}{12\pi^2} 
           \, S_0(m_t^2/m_W^2) \, (V_{tq}^* V_{tb}^{})^2 \, 
\EQN{M_12}          \\
\Gamma_{12} &=& \frac{
           G_F^2 m_b^2 \eta'_B m_{B_q} B_{B_q} f_{B_q}^2}{8\pi} 
           \left[ (V_{tq}^* V_{tb}^{})^2 + 
                   V_{tq}^* V_{tb}^{} V_{cq}^* V_{cb}^{} \,\, 
                         {\cal O}\!\left(\frac{m_c^2}{m_b^2}\right)
                 + (V_{cq}^* V_{cb}^{})^2 \,\,
                         {\cal O}\!\left(\frac{m_c^4}{m_b^4}\right) 
           \right] ~~~ 
\end{eqnarray}
\noindent 
where $G_F$ is the Fermi constant, $m_W$ the $W$ boson mass, 
$m_i$ the mass of quark $i$, 
and $m_{B_q}=M$, $f_{B_q}$ and $B_{B_q}$ are the \Bqq\ mass, 
weak decay constant and bag parameter, respectively.
The known function $S_0(x_t)$ can be approximated very well with 
$0.784\,x_t^{0.76}$\cite{Buras_Fleischer_HeavyFlavorsII}
and $V_{ij}$ are the elements of the CKM matrix\cite{CKM}.
The QCD corrections $\eta_B$ and $\eta'_B$ are of order unity. 
The only non negligible contributions to $M_{12}$ are from box diagrams
involving two top quarks. 
The phases of $M_{12}$ and $\Gamma_{12}$ satisfy
\begin{equation} 
\varphi_{M_{12}} - \varphi_{\Gamma_{12}} = 
\pi + {\cal O}\left(\frac{m^2_c}{m^2_b} \right) 
\EQN{phasediff}
\end{equation} 
implying that the mass eigenstates  
have mass and width differences of opposite signs. This means that,
like in the $K{^0}\hbox{--}\overline K{^0}$ system, the 
``heavy'' state with mass $M_{\rm heavy} = \max(M_+,M_-)$ has a smaller 
decay width than that of the ``light'' state with mass 
$M_{\rm light} = \min(M_+,M_-)$. We thus redefine
\begin{equation} 
\DELTA m = M_{\rm heavy} - M_{\rm light} \,, ~~~ 
\DELTA\Gamma = \Gamma_{\rm light} - \Gamma_{\rm heavy} \,, 
\EQN{redef}
\end{equation} 
where $\DELTA m$ is positive by definition and $\DELTA\Gamma$ is expected to 
be positive in the Standard Model.

Furthermore, the quantity 
\begin{equation} 
\left|\frac{\Gamma_{12}}{M_{12}}\right| \simeq \frac{3\pi}{2}
\frac{m^2_b}{m^2_W} \frac{1}{S_0(m_t^2/m_W^2)} 
\sim {\cal O}\left(\frac{m^2_b}{m^2_t} \right)
\EQN{G12overM12} 
\end{equation} 
is small, and a power expansion of $|q/p|^2$ yields
\begin{equation} 
\left|\frac{q}{p}\right|^2 = 1 + \left|\frac{\Gamma_{12}}{M_{12}}\right| 
\sin(\varphi_{M_{12}}-\varphi_{\Gamma_{12}})
+ {\cal O}\left( \left|\frac{\Gamma_{12}}{M_{12}}\right|^2\right) \,.
\end{equation} 
Therefore, considering both 
\Eqss{phasediff}{G12overM12},
the $CP$-violating parameter
\begin{equation} 
1 - \left|\frac{q}{p}\right|^2 \simeq 
{\rm Im}\left(\frac{\Gamma_{12}}{M_{12}}\right) 
\end{equation} 
is expected to be very small: $\sim {\cal O}(10^{-3})$ for the 
\BdBdbar\ system and $\lsim {\cal O}(10^{-4})$ 
for the \BsBsbar\ system\cite{Bigi}.

In the approximation of negligible $CP$ violation in mixing, 
the ratio $\DELTA\Gamma/\DELTA m$ is equal to the 
small quantity $\left|\Gamma_{12}/M_{12}\right|$ of \Eq{G12overM12}; it is
hence independent of CKM matrix elements, {\it i.e.}
the same for the \BdBdbar\ and \BsBsbar\ systems. It can be calculated 
with lattice QCD techniques; typical results are $\sim 5 \times 10^{-3}$
with quoted uncertainties of $\sim 30\%$.
Given the current experimental knowledge (discussed below) 
on the mixing parameter $x$, 
\begin{equation} 
\cases{
x_d = 0.755 \pm 0.015 & (\BdBdbar\ system) \cr
x_s > 19.0 ~ \hbox{at 95\%~CL} & (\BsBsbar\ system) \cr}
 \,,
\end{equation} 
the Standard Model thus predicts that $\DELTA\Gamma/\Gamma$ is very 
small for the \BdBdbar\ system (below 1\%), 
but considerably larger for the \BsBsbar\ system 
($\sim 10\%$). 
This width difference is caused by the existence of final states to which 
both the \Bqq\ and \Bqbar\ mesons can decay. Such decays involve 
$b \to c\overline{c}q$ quark-level transitions,
which are 
Cabibbo-suppressed if $q=d$ and Cabibbo-allowed if $q=s$.
If the final states common to \Bs\ and \Bsbar\ are 
predominantly $CP$-even as discussed in \Ref{Aleksan},
then the \BsBsbar\ mass eigenstate with the 
largest decay width corresponds to the $CP$-even eigenstate. 
Taking \Eq{redef} into account, one thus expects 
$\Gamma_{\rm light} = \Gamma_+$ and
\begin{equation} 
\DELTA m_s = M_- - M_+ > 0 \,, ~~~ 
\DELTA \Gamma_s = \Gamma_+ - \Gamma_- > 0 \,. 
\end{equation} 

\section*{Experimental issues and methods for oscillation analyses}

Time-integrated measurements of \BBbar\ mixing
were published for the first time in 1987 by UA1\cite{UA1_CP} 
and ARGUS\cite{ARGUS_CP}, 
and since then by many other experiments.
These measurements are typically based on counting same-sign and opposite-sign 
lepton pairs from the semileptonic decay of the produced $b\overline{b}$ pairs.
Such analyses cannot easily separate the contributions from the 
different $b$-hadron species, therefore the clean environment 
of $\Upsilon(4S)$ machines (where only \Bd\ and charged $B_u$ mesons 
are produced) is in principle best suited to measure $\chi_d$.

However, better sensitivity is obtained from time-dependent analyses
aimed at the direct measurement of the oscillation frequencies 
$\DELTA m_d$ and $\DELTA m_s$,
from the proper time
distributions of \Bd\ or \Bs\ candidates 
identified through their decay in (mostly) flavor-specific modes and
suitably tagged as mixed or unmixed.
(This is particularly true for the \BsBsbar\ 
system where the large 
value of $x_s$ implies maximal mixing, {\it i.e.} $\chi_s \simeq 1/2$.)
In such analyses the \Bd\ or \Bs\ mesons are 
either fully reconstructed, partially reconstructed from a charm meson, 
selected from a lepton 
with the characteristics of a $b\to\ell^-$ decay,
or selected from a reconstructed displaced vertex. 
At high-energy colliders (LEP, SLC, Tevatron), 
the proper time $t=\frac{m_B}{p}L$ is measured 
from the distance $L$ between the production vertex and 
the $B$ decay vertex, 
and from an estimate of the $B$ momentum $p$.
At asymmetric $B$ factories (KEKB, PEP-II), producing 
$e^+e^-\to\Upsilon(4S) \to\hbox{\Bd\Bdbar}$ events with a boost
$\beta\gamma$ ($=0.425$, $0.55$),
the proper time difference between the two $B$ candidates
is estimated as $\DELTA t \simeq \frac{\DELTA z}{\beta\gamma c}$, 
where $\DELTA z$ is the spatial separation between the 
two $B$ decay vertices along the boost direction. 
In all cases, the good resolution needed on the vertex positions 
is obtained with silicon detectors. 

The statistical significance ${\cal S}$ 
of a \Bd\ or \Bs\ oscillation signal can be approximated as\cite{amplitude}
\IndexPageno{Bstatsigm}
\begin{equation} 
{\cal S} \approx \sqrt{N/2} \,f_{\rm sig}\, (1-2\eta)\,
e^{-(\DELTA m\,\sigma_t)^2/2}  \,, 
\EQN{significance} 
\end{equation} 
where $N$ and $f_{\rm sig}$
are the number of candidates and the fraction of signal
in the selected sample, $\eta$ is the total mistag probability, 
and $\sigma_t$ is the resolution on proper time (or proper time difference). 
The quantity ${\cal S}$ decreases very quickly as 
$\DELTA m$ increases; this dependence is controlled by $\sigma_t$,
which is therefore a critical parameter for $\DELTA m_s$ analyses. 
At high-energy colliders, the proper time resolution 
$\sigma_t \sim \frac{m_B}{\langle p\rangle} \sigma_L 
\oplus t \frac{\sigma_p}{p}$ 
includes a constant contribution due to the decay length resolution 
$\sigma_L$ (typically 0.05--0.3~ps), and a term due to the 
relative momentum resolution $\sigma_p/p$ (typically 10--20\%
for partially reconstructed decays)  
which increases with proper time.
At $B$ factories, the $B$ momentum is reconstructed and/or 
estimated from the beam energy constraint, and the 
term due to the spatial resolution dominates
(typically 1--1.5~ps because of the much smaller $B$ boost).

In order to tag a $B$ candidate 
as mixed or unmixed, it is necessary
to determine its flavor 
both in the initial state and in the final state. 
The initial and final state mistag probabilities,\IndexPageno{Bmistagm}
$\eta_i$ and $\eta_f$, degrade ${\cal S}$
by a total factor $(1-2\eta)=(1-2\eta_i)(1-2\eta_f)$.
In lepton-based analyses, the final state is tagged by the charge of 
the lepton from $b\to\ell^-$ decays; the biggest contribution to $\eta_f$ 
is then due to $\overline{b}\to\overline{c}\to\ell^-$ decays. 
Alternatively, the charge of a 
reconstructed charm meson ($D^{*-}$ from \Bd\ or $D_s^-$ from \Bs), 
or that of a kaon thought to come from a $b\to c\to s$
decay\cite{SLD_dmd_prelim}, can be used.
For fully inclusive analyses based on topological 
vertexing, final state tagging techniques include 
jet charge\cite{ALEPH_dmd_prelim} and charge 
dipole\cite{SLD_dms_prelim_Thom}\cite{DELPHI_dmd_dms_prelim} methods.

At high-energy colliders, the methods to tag the initial state 
(i.e.\ the state at production), 
can be divided in two groups: the ones 
that tag the initial charge of the $\overline{b}$ quark contained in the 
$B$ candidate itself (same-side tag),\IndexPageno{bmixam}
 and the ones that tag the initial 
charge of the other $b$ quark produced in the event (opposite-side tag). 
On the same side, the charge of a track from the primary vertex is 
correlated with the production state of the $B$ if that track is a decay
product of a $B^{**}$ state or the first particle in the fragmentation 
chain\cite{CDF_dmd}\cite{ALEPH_dms}.
Jet- and vertex-charge techniques work on both sides and on the opposite
side, respectively. 
Finally, the charge of a lepton from $b\to\ell^-$ or of a kaon
from $b\to c\to s$ can be used as opposite side tags,
keeping in mind that their performance is degraded due to integrated mixing.
At SLC, the beam polarization produced a sizeable forward-backward 
asymmetry in the $Z\to b\overline{b}$ decays and provided another
very interesting and effective initial state tag based on the polar angle 
of the $B$ candidate\cite{SLD_dmd_prelim}.
Initial state tags have also been combined to 
reach $\eta_i \sim 26\%$ at LEP\cite{ALEPH_dms}\cite{DELPHI_dms_dgs},
or even 22\% at SLD\cite{SLD_dms_prelim_Thom} with full efficiency. 
The equivalent figure at CDF (Tevatron Run I) 
is $\sim 40\%$\cite{Paulini}.

At $B$ factories, the flavor of a \Bd\ meson at production cannot 
be determined, since the two neutral $B$ mesons produced in a 
$\Upsilon(4S)$ decay evolve in a coherent $P$-wave state where they 
keep opposite flavors at any time.
However, as soon as one of them decays, the other follows a time-evolution
given by \Eqssor{time_evol1}{time_evol2}, 
where $t$ is replaced with $\DELTA t$. Hence, the ``initial state''
tag of a $B$ can be taken as the final state tag of the other $B$. 
Effective mistag probabilities of $\eta_i \sim 24 \%$ 
for full efficiency (corresponding to effective tagging efficiencies of 
$ \sim 27 \%$ for perfect tagging) are achieved by BABAR 
and Belle\cite{sin2beta}, 
using different techniques including $b \to \ell^-$ and 
$b\to c\to s$ tags. 
It is interesting to note that, in this case, 
mixing of this other $B$ (i.e.\ the coherent mixing occurring before
the first $B$ decay) does not contribute to the mistag probability.

In the absence of experimental evidence for a width difference,
oscillation analyses typically neglect $\DELTA\Gamma$
and describe the data with the physics functions
$\Gamma e^{-\Gamma t} (1 \pm \cos( \DELTA m t))/2$
(high-energy colliders) or
$\Gamma e^{-\Gamma |\DELTA t|} (1 \pm \cos( \DELTA m \DELTA t))/4$
(asymmetric $\Upsilon(4S)$ machines).
As can be seen from \Eq{cosh_cos}, a non-zero value of $\DELTA\Gamma$
would effectively reduce the oscillation amplitude with a small 
time-dependent factor that would be very difficult to distinguish 
from time resolution effects. 
Whereas measurements of $\DELTA m_d$ are usually extracted from the data
using a maximum likelihood fit, no significant \BsBsbar\ 
oscillations have been seen so far.
To extract information useful to set lower limits on $\DELTA m_s$,
\Bs\ analyses follow a method\cite{amplitude}
in which a \Bs\ oscillation amplitude ${\cal A}$
is measured as a function of a fixed test value of $\DELTA m_s$, 
using a maximum likelihood fit based on the functions
$\Gamma_s e^{-\Gamma_s t} (1 \pm {\cal A} \cos( \DELTA m_s t))/2$. 
To a very good approximation, the statistical uncertainty on ${\cal A}$
is Gaussian and equal to $1/{\cal S}$\cite{amplitude}.
If $\DELTA m_s=\DELTA m_s^{\rm true}$, one expects
${\cal A} = 1 $ within the total uncertainty $\sigma_{\cal A}$;
however, if $\DELTA m_s$ is (far) below its
true value, a measurement consistent with ${\cal A} = 0$ is expected.
A value of $\DELTA m_s$ can be excluded at 95\%~CL
if ${\cal A} + 1.645\,\sigma_{\cal A} \le 1$.
If $\DELTA m_s^{\rm true}$ is very large, one expects ${\cal A} = 0$,
and all values of $\DELTA m_s$ such that $1.645\, \sigma_{\cal A}(\DELTA m_s) < 1$
are expected to be excluded at 95\%~CL.
Because of the proper time resolution, the quantity $\sigma_{\cal A}(\DELTA m_s)$
is an increasing function of $\DELTA m_s$ and one therefore
expects to be able to exclude individual
$\DELTA m_s$ values up to $\DELTA m_s^{\rm sens}$, 
where $\DELTA m_s^{\rm sens}$, called here the
sensitivity of the analysis, is defined by
$1.645\,\sigma_{\cal A}(\DELTA m_s^{\rm sens}) =1$. 

\section*{\boldmath \Bd\ mixing studies}


Many \BdBdbar\ oscillations analyses have been 
performed by the ALEPH\cite{ALEPH_dmd_prelim}\cite{ALEPH_dmd}, 
BABAR\cite{BABAR_dmd}, Belle\cite{Belle_dmd}\cite{Belle_dmd_prelim},
CDF\cite{CDF_dmd}\cite{CDF_dmd_prelim}, 
DELPHI\cite{DELPHI_dmd_dms_prelim}\cite{DELPHI_dmd}, 
L3\cite{L3_dmd}, OPAL\cite{OPAL_dmd} and SLD\cite{SLD_dmd_prelim} 
collaborations.
Although a variety of different techniques have been used, the 
individual $\DELTA m_d$ 
results obtained at high-energy colliders have remarkably similar precision.
Their average is compatible with the recent and more precise measurements 
from asymmetric $B$ factories.
The systematic uncertainties are not negligible; 
they are often dominated by sample composition, mistag probability,
or $b$-hadron lifetime contributions.
Before being combined, the measurements are adjusted on the basis of a 
common set of input values, including the $b$-hadron lifetimes and fractions
published in this {\it Review}. Some measurements are statistically correlated. 
Systematic correlations arise both from common physics sources (fragmentation 
fractions, lifetimes, branching ratios of $b$~hadrons), and from purely 
experimental or algorithmic effects (efficiency, resolution, tagging, 
background description). Combining all published measurements%
\cite{CDF_dmd,ALEPH_dmd,BABAR_dmd,Belle_dmd,DELPHI_dmd,L3_dmd,OPAL_dmd} 
and accounting for all identified correlations 
as described in \Ref{HF_preprint} yields 
$\DELTA m_d = {\rm 0.489 \pm 0.005 (stat) \pm 0.007 (syst)}~\hbox{ps}^{-1}$.

On the other hand, ARGUS and CLEO have published time-integrated 
measurements\cite{ARGUS_chid,CLEO_chid_CP,CLEO_chid_CP_y}, 
which average to $\chi_d = 0.182 \pm 0.015$.
Following \Ref{CLEO_chid_CP_y}, 
the width difference $\DELTA \Gamma_d$ could 
in principle be extracted from the
measured value of $\Gamma_d$ and the above averages for 
$\DELTA m_d$ and $\chi_d$ 
(see \Eqss{x}{chi}), 
provided that $\DELTA \Gamma_d$ has a negligible impact on 
the $\DELTA m_d$ measurements. However, a stronger 
constraint, $\DELTA \Gamma_d/\Gamma_d < 20\%$ at 90\% CL, 
has been obtained by DELPHI from a direct 
time-dependent study\cite{DELPHI_dmd_dms_prelim}.
Assuming $\DELTA \Gamma_d =0$ and no CP violation in mixing, 
and using the measured \Bd\ lifetime,
the $\DELTA m_d$ and $\chi_d$ results are combined to yield the 
world average
\begin{equation} 
\DELTA m_d = 0.489 \pm 0.008~\hbox{ps}^{-1} 
\EQN{dmdw}
\end{equation} 
or, equivalently,
\begin{equation} 
\chi_d=0.181\pm 0.004\,.  
\EQN{chidw}
\end{equation} 


Evidence for $CP$ violation in \Bd\ mixing has been searched for,
both with flavor-specific and inclusive \Bd\ decays, 
in samples where the initial 
flavor state is tagged. In the case of semileptonic 
(or other flavor-specific) decays, 
where the final state tag is 
also available, the following asymmetry
\begin{equation}
\frac{
N(\hbox{\Bdbar}(t) \to \ell^+           \nu_{\ell} X) -
N(\hbox{\Bd}(t)    \to \ell^- \overline{\nu}_{\ell} X) }{
N(\hbox{\Bdbar}(t) \to \ell^+           \nu_{\ell} X) +
N(\hbox{\Bd}(t)    \to \ell^- \overline{\nu}_{\ell} X) } 
= a_{CP} \simeq 1 - |q/p|^2_d 
\simeq \frac{4{\rm Re}(\epsilon_d)}{1+|\epsilon_d|^2} 
\end{equation}
has been measured, either in time-integrated analyses at 
CLEO\cite{CLEO_chid_CP,CLEO_chid_CP_y,CLEO_CP_semi} 
and CDF\cite{CDF_CP_semi}, or in time-dependent analyses at 
LEP\cite{OPAL_CP_semi,DELPHI_CP,ALEPH_CP} and 
BABAR\cite{BABAR_CP_semi}.
In the inclusive case, also investigated at 
LEP\cite{DELPHI_CP}\cite{ALEPH_CP}\cite{OPAL_CP_incl},
no final state tag is used, and the asymmetry\cite{incl_asym}
\begin{equation}
\frac{
N(\hbox{\Bd}(t) \to {\rm all}) -
N(\hbox{\Bdbar}(t) \to {\rm all}) }{
N(\hbox{\Bd}(t) \to {\rm all}) +
N(\hbox{\Bdbar}(t) \to {\rm all}) } 
\simeq
a_{CP} \left[ \frac{x_d}{2} \sin(\DELTA m_d \,t) - 
\sin^2\left(\frac{\DELTA m_d \,t}{2}\right)\right] 
\end{equation}
must be measured as a function of the proper time to extract information 
on $CP$ violation.
In all cases asymmetries compatible with zero have been found,  
with a precision limited by the available statistics. A simple 
average of all published and preliminary 
results\cite{CLEO_chid_CP,CLEO_chid_CP_y,CLEO_CP_semi,CDF_CP_semi,%
OPAL_CP_semi,DELPHI_CP,ALEPH_CP,BABAR_CP_semi,OPAL_CP_incl}
neglecting small possible statistical correlations and 
assuming half of the systematics to be correlated between measurements 
performed at the same energy, is
$a_{CP} = {\rm -0.002 \pm 0.009(stat) \pm 0.008(syst)}$,
a result which does not yet constrain the Standard Model.

The $\DELTA m_d$ result of \Eq{dmdw} provides an estimate of $|M_{12}|$ and 
can be used, together with \Eqss{Dm,DG}{M_12}, 
to extract the magnitude of the CKM matrix element $V_{td}$ 
within the Standard Model\cite{CKM_review}. The main experimental 
uncertainties on the resulting estimate of $|V_{td}|$ come from 
$m_t$ and $\DELTA m_d$; however, the extraction is at present 
completely dominated by the uncertainty on the hadronic 
matrix element $f_{B_d} \sqrt{B_{B_d}} = 230\pm 40$~MeV 
obtained from lattice QCD calculations\cite{lattice_QCD}.

\section*{\boldmath \Bs\ mixing studies}

\BsBsbar\ oscillations have been the subject of many 
studies from ALEPH\cite{ALEPH_dms}\cite{ALEPH_dms_prelim}, CDF\cite{CDF_dms}, 
DELPHI\cite{DELPHI_dmd_dms_prelim}\cite{DELPHI_dms_dgs}\cite{DELPHI_dms}\cite{DELPHI_dms_prelim},
OPAL\cite{OPAL_dms} and SLD\cite{SLD_dms_prelim_Thom}\cite{SLD_dms_prelim}. 
No oscillation signal has been found so far. 
The most sensitive analyses appear to be the ones based 
on inclusive lepton samples at LEP. Because of their better 
proper time resolution, the small data samples analyzed 
inclusively at SLD, as well as the few fully reconstructed $B_s$ decays 
at LEP, turn out to be also very useful to explore the high $\DELTA m_s$  
region.

All results are limited by the available statistics.
They can easily be combined, since all experiments provide 
measurements of the \Bs\ oscillation amplitude. The latter are averaged 
using the procedure of \Ref{HF_preprint} to yield the combined amplitudes 
${\cal A}$ shown in \Fig{amplitude} as a function of $\DELTA m_s$. 
The individual results 
have been adjusted to common physics inputs, and all known correlations 
have been accounted for; 
the sensitivities of the inclusive analyses, 
which depend directly through \Eq{significance} 
on the assumed fraction $f_s$
of \Bs\ mesons in an unbiased sample of weakly-decaying $b$~hadrons, 
have also been rescaled to a common preliminary average of 
$f_s = 0.097 \pm 0.011$.
The combined sensitivity for 95\%~CL exclusion of $\DELTA m_s$ values is found 
to be 19.3~ps$^{-1}$.
All values of $\DELTA m_s$ below 14.9~ps$^{-1}$ are excluded at 95\%~CL.
The values between 14.9 and 22.4~ps$^{-1}$ cannot be excluded, because
the data is compatible with a signal in this region. However,
no deviation from ${\cal A}=0$ is seen in \Fig{amplitude} that would
indicate the observation of a signal.

\begin{figure}\begin{center}
\vskip -2cm 
    \parindent = 0pt \leftskip = 0pt \rightskip = 0pt%
    \vskip .4in%
    \leavevmode%
    \centerline{%
\psfig{figure=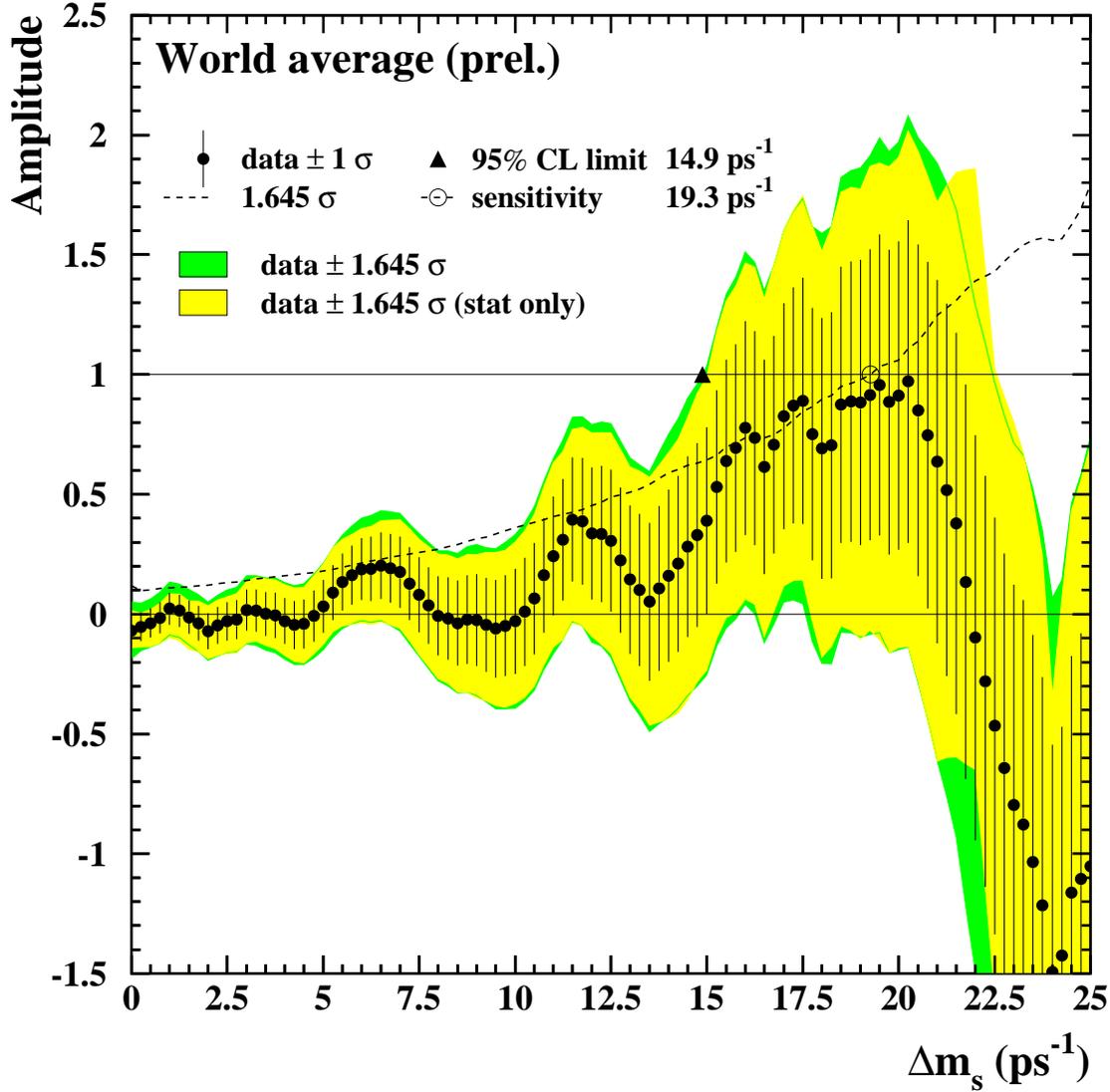,width=\textwidth,clip=t}}%
    \nobreak%
    \vglue .1in%
    \nobreak%
\vglue -0.5cm
\caption{
Combined measurements of the \Bs\ oscillation amplitude as a 
function of $\DELTA m_s$, including all preliminary results 
available at the time of the 
winter 2002 conferences\cite{winter_2002_abbaneo}. 
The measurements are dominated by statistical uncertainties. 
Neighboring points are statistically correlated.
}
\labf{amplitude}
\end{center}\end{figure}

Some $\DELTA m_s$ analyses are still 
unpublished\cite{SLD_dms_prelim_Thom}\cite{DELPHI_dmd_dms_prelim}\cite{ALEPH_dms_prelim}\cite{DELPHI_dms_prelim}\cite{SLD_dms_prelim}.
Using only published results, the combined $\DELTA m_s$ result is 
\begin{equation} 
\DELTA m_s > 13.1~\hbox{ps}^{-1} ~~~ \hbox{at 95\%~CL} \,,
\end{equation} 
with a sensitivity of 13.3~ps$^{-1}$. 

The information on $|V_{ts}|$ obtained, 
in the framework of the Standard Model, from the combined amplitude spectrum 
is hampered by the hadronic uncertainty, as in the \Bd\ case. 
However, many uncertainties cancel in the frequency ratio
\begin{equation} 
\frac{\DELTA m_s}{\DELTA m_d} = \frac{m_{B_s}}{m_{B_d}}\, \xi^2
                    \left|\frac{V_{ts}}{V_{td}}\right|^2 \,,
\EQN{ratio} 
\end{equation} 
where $\xi= (f_{B_s} \sqrt{B_{B_s}})/(f_{B_d} \sqrt{B_{B_d}})=1.16\pm 0.05$
is an SU(3) flavor-symmetry breaking factor 
obtained from lattice QCD calculations\cite{lattice_QCD}.
The CKM matrix can be constrained using the experimental results 
on $\DELTA m_d$, 
$\DELTA m_s$, $|V_{ub}/V_{cb}|$, $\epsilon_K$ and $\sin(2\beta)$ 
together with theoretical inputs and unitarity 
conditions\cite{CKM_review}\cite{CKM_workshop}.
Given all
measurements other than $\DELTA m_d$ and $\DELTA m_s$, 
the constraint from our knowledge on the ratio $\DELTA m_d/\DELTA m_s$
is presently more effective in limiting the position of the apex of the 
CKM unitarity triangle than the one obtained from the $\DELTA m_d$ 
measurements alone, due to the reduced hadronic uncertainty in \Eq{ratio}.
We note also that it would be difficult for the Standard Model to accommodate 
values of $\DELTA m_s$ above $\sim 25~{\rm ps}^{-1}$\cite{CKM_workshop}.

Information on $\DELTA\Gamma_s$ can be obtained by studying the proper time 
distribution of untagged data samples enriched in 
\Bs\ mesons\cite{Hartkorn_Moser}.
In the case of an inclusive \Bs\ selection\cite{L3_DGs} or a semileptonic 
\Bs\ decay selection\cite{DELPHI_dms_dgs}\cite{CDF_DGs}, 
both the short- and long-lived
components are present, and the proper time distribution is a superposition 
of two exponentials with decay constants 
$\Gamma_s\pm \DELTA\Gamma_s/2$.
In principle, this provides sensitivity to both $\Gamma_s$ and 
$(\DELTA\Gamma_s/\Gamma_s)^2$. Ignoring $\DELTA\Gamma_s$ and fitting for 
a single exponential leads to an estimate of $\Gamma_s$ with a 
relative bias proportional to $(\DELTA\Gamma_s/\Gamma_s)^2$. 
An alternative approach, which is directly sensitive to first order in 
$\DELTA\Gamma_s/\Gamma_s$, 
is to determine the lifetime of \Bs\ candidates decaying to $CP$ 
eigenstates; measurements exist for 
$\hbox{\Bs} \to J/\psi \phi$\cite{CDF_Jpsiphi} and 
$\hbox{\Bs} \to D_s^{(*)+} D_s^{(*)-}$\cite{ALEPH_DGs}, which are 
mostly $CP$-even states\cite{Aleksan}. An estimate of $\DELTA\Gamma_s/\Gamma_s$
has also been obtained directly from a measurement of the 
$\hbox{\Bs} \to D_s^{(*)+} D_s^{(*)-}$ branching ratio\cite{ALEPH_DGs}, 
under the assumption that 
these decays practically account for all the $CP$-even final states. 

Present data is not precise enough to efficiently constrain 
both $\Gamma_s$ and $\DELTA\Gamma_s/\Gamma_s$; since the \Bs\ and \Bd\ 
lifetimes are predicted to be equal within less than a 
percent\cite{equal_lifetimes}, an expectation compatible with 
the current experimental data\cite{B_decays},
the constraint $\Gamma_s = \Gamma_d$ can also be used to improve the
extraction of $\DELTA \Gamma_s/\Gamma_s$.
Applying the combination procedure of \Ref{HF_preprint} 
on the published results\cite{DELPHI_dms_dgs}%
\cite{CDF_DGs,CDF_Jpsiphi,ALEPH_DGs,ALEPH_OPAL_Dsl_lifetime}
yields
\begin{equation} 
\DELTA\Gamma_s/\Gamma_s < 0.52 ~~~ \hbox{at 95\%~CL}
\end{equation} 
without external constraint, or
\begin{equation} 
\DELTA\Gamma_s/\Gamma_s < 0.31 ~~~ \hbox{at 95\%~CL}  
\end{equation} 
when constraining $1/\Gamma_s$ to the measured \Bd\ lifetime.
These results are not yet precise enough to test Standard Model predictions.

\section*{\boldmath  Average $b$-hadron mixing and 
$b$-hadron production fractions at high energy}

Let $f_u$, $f_d$, $f_s$ and $f_{\rm baryon}$ 
be the $B_u$, \Bd, \Bs\ and $b$-baryon fractions composing an 
unbiased sample of weakly-decaying $b$~hadrons
produced in high-energy colliders. 
LEP experiments have measured 
$f_s \times {\rm BR}(B^0_s \to D_s^- \ell^+ \nu_\ell X)$\cite{LEP_fs}, 
${\rm BR}(b \to \Lambda_b^0) \times 
{\rm BR}(\Lambda_b^0 \to \Lambda_c^+\ell^- \overline\nu_\ell X)$\cite{LEP_fla}
and ${\rm BR}(b \to \Xi_b^-) \times 
{\rm BR}(\Xi_b^- \to \Xi^-\ell^-\overline\nu_\ell X)$\cite{LEP_fxi}
from partially reconstructed final states 
including a lepton, $f_{\rm baryon}$ 
from protons identified in $b$ events\cite{ALEPH-fbar}, and the 
production rate of charged $b$ hadrons\cite{DELPHI-fch}. 
The various $b$-hadron fractions 
have also been measured at CDF from electron-charm final states\cite{CDF_f}.
All the published results have been combined 
following the procedure and assumptions described in \Ref{HF_preprint}, 
to yield
$f_u = f_d = (37.3\pm 2.0)\%$, 
$f_s = (13.9\pm 3.8)\%$ and $f_{\rm baryon}=(11.5 \pm 2.0)\%$ under the 
constraints
\begin{equation} 
f_u = f_d ~~~~\hbox{and}~~~ f_u + f_d + f_s + f_{\rm baryon} = 1 \,. 
\EQN{constraints}
\end{equation} 

Time-integrated mixing analyses performed with lepton pairs 
from $b\overline{b}$ 
events produced at high-energy colliders measure the quantity 
\begin{equation} 
\overline{\chi} = f'_d \,\chi_d + f'_s \,\chi_s \,,
\end{equation} 
where $f'_d$ and $f'_s$ are the fractions of \Bd\ and \Bs\ hadrons 
in a sample of semileptonic $b$-hadron decays. 
Assuming that all $b$~hadrons have the same semileptonic decay width implies 
$f'_q = f_q/(\Gamma_q \tau_b)$ ($q=s,d$), where 
$\tau_b$ is the average $b$-hadron lifetime. 
Hence $\overline{\chi}$ measurements can be used to improve our knowledge on 
the fractions 
$f_u$, $f_d$, $f_s$ and $f_{\rm baryon}$.

Combining the above estimates of these fractions with 
the average $\overline{\chi} = 0.1184 \pm 0.0045$ 
(published in this {\it Review}),
$\chi_d$ from \Eq{chidw} and
$\chi_s = 1/2$ yields, under the constraints of \Eq{constraints},
\begin{eqnarray}
f_u = f_d &=& (38.8 \pm 1.3)\% \,, 
\\
f_s &=& (10.6 \pm 1.3)\% \,, 
\\
f_{\rm baryon} &=& (11.8 \pm 2.0)\% \,,
\end{eqnarray}
showing that mixing information substantially reduces the uncertainty on 
$f_s$. These results and the averages quoted in 
\Eqss{dmdw}{chidw}
for $\chi_d$ and $\DELTA m_d$ have been obtained in a consistent way
by the $B$ oscillations working group\cite{HF_preprint}, 
taking into account the fact that many individual measurements of 
$\DELTA m_d$ depend on the assumed values for the $b$-hadron fractions. 


\section*{Summary and prospects}

\BBbar\ mixing has been and still is a field of intense study.
The mass difference in the \BdBdbar\ system is very well measured 
(with an accuracy of $1.7\%$) but, 
despite an impressive theoretical effort, 
the hadronic uncertainty still limits the precision of the 
extracted estimate of $|V_{td}|$. 
The mass difference in the \BsBsbar\
system is much larger and still unmeasured. 
However, the current experimental lower limit on $\DELTA m_s$ already provides, 
together with $\DELTA m_d$, a significant constraint on the CKM matrix within the
Standard Model. 
No strong experimental evidence exists yet for the rather large decay width 
difference expected in the \BsBsbar\
system. It is interesting to recall that 
the ratio $\DELTA \Gamma_s/\DELTA m_s$ does not depend on CKM matrix elements 
in the Standard Model
(see \Eq{G12overM12}), and that a measurement of either $\DELTA m_s$ 
or $\DELTA \Gamma_s$ could be turned into a Standard Model 
prediction of the other one. 

The LEP and SLD experiments have still not finalized all their 
\Bs\ oscillation analyses, but a first measurement of $\DELTA m_s$ 
from data collected at the $Z$ pole is now very unlikely. 
In the near future, the most 
promising prospects for \Bs\ mixing are from Run II at the Tevatron, 
where both $\DELTA m_s$ and $\DELTA \Gamma_s$ are expected to be 
measured with fully reconstructed \Bs\ decays;
for example, with $2~\hbox{fb}^{-1}$ of data, 
CDF expects to observe \Bs\ oscillations for values of $\DELTA m_s$ 
up to $\sim 40-50~\hbox{ps}^{-1}$ (depending on event yields 
and signal-to-background ratios)\cite{TeV_RunII}, well above the current 
Standard Model prediction.

$CP$ violation in $B$ mixing, which has not been seen yet, as well as the 
phases involved in $B$ mixing, will be further investigated with the large 
statistics that will become available both at the $B$ factories and at the 
Tevatron.

$B$ mixing may not have delivered all its secrets yet, because 
it is one of the phenomena where new physics might very well reveal itself 
(for example new particles involved in the box diagrams). 
Theoretical calculations in lattice QCD are becoming more reliable and 
further progress in reducing hadronic uncertainties is expected. 
In the long term, a stringent check of the consistency, within the 
Standard Model, of the \Bd\ and 
\Bs\ mixing measurements with all other measured observables in $B$ physics
(including $CP$ asymmetries in $B$ decays) will be possible, 
allowing to place limits on new physics or, better, discover new physics.

\end{document}